\documentclass[superscriptaddress,twocolumn,showpacs,amsmath,amssymb]{revtex4}

\usepackage{amsfonts}
\usepackage{amsmath}
\usepackage{amssymb}
\usepackage{graphicx}
\usepackage{rotating}
\usepackage{tabularx}%
\usepackage{color}
\begin{document}

\affiliation{UMR 8608, Institut de Physique Nucl\'eaire, 
91406 Orsay Cedex, France
}

\title{On the Wong cross section and fusion oscillations}
\author{N. Rowley}
\affiliation{UMR 8608, Institut de Physique Nucl\'eaire, 
91406 Orsay Cedex, France
}

\author{K. Hagino}
\affiliation{Department of Physics, Tohoku University, Sendai 980-8578, Japan}
\affiliation{Research Center for Electron Photon Science, Tohoku University,
1-2-1 Mikamine, Sendai 982-0826, Japan}


\begin{abstract}
We re-examine the well-known Wong formula for heavy-ion fusion cross sections.
Although this celebrated 
formula yields almost exact results for
single-channel calculations for relatively heavy systems
such as $^{16}$O+$^{144}$Sm, 
it tends to overestimate the cross section for light systems such
as $^{12}$C+$^{12}$C. 
We generalise the formula to take account of the
energy dependence of the barrier parameters and show that the
energy-dependent version
gives results practically indistinguishable from a full quantal calculation.
We then examine the deviations arising from the discrete nature 
of the intervening angular momenta, whose effect can lead to an oscillatory contribution
to the excitation function.
We recall some compact, analytic expressions for these oscillations,
and highlight the important physical parameters that give rise to them.
Oscillations in symmetric systems are discussed,
as are systems where the target and projectile identities can be exchanged 
via a strong transfer channel.  
\end{abstract}


\maketitle

\section{Introduction}

It has been known for many years that fusion cross sections for 
several light heavy-ion systems, such as 
$^{12}$C+$^{12}$C (Ref.~\cite{12c12c}), $^{12}$C+$^{16}$O (Ref.~\cite{12c16o})
and $^{16}$O+$^{16}$O (Ref.~\cite{16o16o}), exhibit 
an oscillatory structure as a function of the incident energy. 
Two recent papers by Esbensen~\cite{esbensen} and Wong~\cite{wong-osc}
interpret the oscillations 
as due to the addition of successive
individual partial waves as the energy increases. 
The effect is of course 
most important for identical spin-zero nuclei, because in that case the
odd partial waves are totally absent and the relevant energy spacing between
successive contributing angular momenta is consequently much larger. 
Both of these authors discuss this phenomenon in the context of the well-known 
Wong fusion cross section~\cite{wong1} derived from the 
Hill-Wheeler expression~\cite{hill-wheeler} for the penetration of a parabolic potential
barrier. 
See also Ref.~\cite{simenel} for a recent publication in a similar context.

In fact the above interpretation was first proposed some 30 years ago by 
Poff\'e, Rowley and Lindsay~\cite{poffe} who, furthermore, gave a
compact and accurate analytic expression for the oscillations that displays
succinctly the dependence on the relevant physical parameters of
the system.
Our purpose in this paper is to discuss the derivation of the Wong cross
section --
both its smooth and oscillatory terms -- and to
present some inadequacies of the 
standard fusion formula that treats the barrier height, position and curvature
$[B,R_B,\hbar\omega]$ as independent of the incident energy $E$ and
thus, implicitly, independent of the angular momentum $l$. 

This re-analysis of the qualities and weaknesses of the Wong formula is
important
in order to be able to distinguish between discrepancies arising from the
approximations used in its derivation and those arising from additional physical
effects such as entrance-channel couplings or limitations on 
compound-nucleus formation.

The paper is organized as follows. In Sec.~II, we derive the Wong
cross section and propose an extended formula which takes into
account the energy dependence of the barrier parameters. We show that the
effect of the energy dependence is significant, particularly for light
heavy-ion systems, and that
the generalised formula reproduces well a full quantal
calculation. In Sec.~III, we discuss the oscillatory part of fusion cross
sections. We re-analyze the compact formula for these oscillation using
the energy-dependent version of the Wong formula. We discuss the fusion of two
identical nuclei as well as fusion between similar nuclei. 
Fusion oscillations in heavier systems are also discussed.
In Sec.~IV, we present our
analyses of the experimental data for the $^{12}$C+$^{12}$C and
$^{12}$C+$^{13}$C systems.  
We summarize the paper in Sec.~V.

\section{Wong formula for fusion cross sections and its improvement}

In this section, 
we compare the results of the Wong expression for the fusion cross
section that contains the three parameters $[B,R_B,\hbar\omega]$
(see below) and the
fusion cross section coming from a quantal calculation with a
specified real potential 
and an absorption that is essentially {\it black box}.
This can be achieved either with
an ingoing-wave boundary condition or with an appropriately
chosen imaginary potential.
Here we choose the latter method, checking carefully that our results are 
insensitive to changes in the imaginary potential used.
With a potential model,
the barrier position $R_B$, its height $B$ and
its curvature $\hbar\omega$ are all 
fixed by the potential parameters.
Indeed it is a very general result that 
the nuclear potential is essentially exponential in the tail,
the region in which 
the Coulomb barrier occurs, at least for relatively light systems with
low $Z_1Z_2$.
In that case, all three of the above parameters are determined by
two potential
parameters, the depth $V_0$ and the surface diffuseness $a$. 

We choose to fix these parameters in the following, much more transparent
way: for the $l=0$ barrier we have
\begin{eqnarray}
  \left.\frac{dV}{dr}\right|_{r=R_B}=
\left.\frac{dV_N}{dr}\right|_{r=R_B}+\left.\frac{dV_C}{dr}\right|_{r=R_B}
=0,
\end{eqnarray}
where  $R_B$ is the position of the barrier. 
Writing $V_N(r)=V_0\,\exp[-(r-R_B)/a]$, we then find
\begin{eqnarray}
\frac{Z_1Z_2e^2}{R_B^2}-\frac{1}{a}\,V_0=0,
\end{eqnarray}
thus obtaining
\begin{eqnarray}
V_N(r)=-\frac{a\,Z_1Z_2e^2}{R_B^2}\,\exp\left(-\frac{r-R_B}{a}\right).
\label{eq:exp-pot}
\end{eqnarray}
Furthermore, we then have a Coulomb plus nuclear barrier
\begin{eqnarray}
B=V(R_B)=\frac{Z_1Z_2e^2}{R_B}\left(1- \frac{a}{R_B}\right),
\label{eq:exp-B}
\end{eqnarray}
and this quadratic equation for $R_B$ gives the barrier position as
\begin{eqnarray}
R_B=\frac{1}{2}\,R_C\left(1+\sqrt{1-4\frac{a}{R_C}}\right),
\label{eq:rb}
\end{eqnarray}
where $R_C$ is just $Z_1Z_2e^2/B$~\footnote{For $a\ll R_C$ this gives the
very simple relation $R_B\approx R_C-a$,
though we shall not use this approximation in this paper.}.
We see that once the barrier height $B$ is given, the position
depends only on the surface diffuseness of the potential, and this also
fixes the strength of the nuclear potential. This is the reason
that it is generally
sufficient to quote the parameters $[B,a]$ in our coupled-channels
({\tt CCFULL}~\cite{ccfull}) 
calculations.
It is of course also sufficient for simpler optical-model/classical
calculations of the fusion cross section.
This procedure is useful because the parameter best determined
by the fusion data is $B$, and furthermore there are various
prescriptions for obtaining
a good theoretical value for this quantity; for example the
Bass~\cite{bass} and the 
Aky\"uz-Winther~\cite{akyuz} potentials that give very similar
$B$ values over a
wide range of heavy-ion systems,
both yielding $B=6.1$~MeV for the $^{12}$C+$^{12}$C
system mainly discussed in this paper. 

As noted above, this reduces the three independent parameters of the Wong 
expression to two, thereby providing a much more rigorous constraint on the
physics of the problem, as we shall see below.
Notice that the exponential potential tends to be deep, and 
such deep potentials have been advocated in Refs.~\cite{RDM77,KRS89,KMR90}. 

If all of the flux crossing the Coulomb barrier fuses,
then the fusion cross section 
is given by
\begin{eqnarray}
\sigma=\frac{\pi\hbar^2}{2mE}\sum_{l=0}^\infty (2l+1)T_l, 
\label{eq:sum}
\end{eqnarray}
and using the Hill-Wheeler formula for the transmission $T_l$
through a parabolic barrier,
we have~\cite{wong1}
\begin{eqnarray}
\sigma=\frac{\pi\hbar^2}{2mE}\sum_{l=0}^\infty \frac{2l+1}{1+\exp
\left[\frac{2\pi}{\hbar\omega}\left(B+\frac{l(l+1)\hbar^2}{2mR_B^2} -E\right)\right]},
\label{eq:hill}
\end{eqnarray}
where in the barrier region, the potential is taken as
$V(r)=B+\frac{l(l+1)\hbar^2}{2mR_B^2}
-\frac{1}{2}m\omega^2(r-R_B)^2$.
(The quantity $\hbar\omega$ is the quantum of energy corresponding
to the inverted barrier and
is generally referred to as the `barrier curvature' [see Eq.~(\ref{eq:vpp})].)
Taking $[B,R_B,\hbar\omega]$ as fixed, and replacing the summation by
an integral, one
obtains the very influential Wong formula 
for the fusion cross section for a single potential barrier
\begin{eqnarray}
  \sigma=\frac{\hbar\omega}{2E}R_B^2\,\ln\left(1+\exp\left[
    \frac{2\pi}{\hbar\omega}(E-B)\right]\right),
\label{eq:wong0}
\end{eqnarray}
which yields, in the limit $E-B\gg\hbar\omega/2\pi$, the classical result
\begin{eqnarray}
E\sigma=\pi R_B^2\,(E-B).
\label{eq:fus_cl}
\end{eqnarray}
The first and second derivatives of this classical equation and its quantal version
yield
\begin{eqnarray}
\frac{d(E\sigma)}{dE}=\pi R_B^2\,\theta(E-B)\rightarrow \pi R_B^2\,
\frac{1}{1+e^x}
\end{eqnarray}
and 
\begin{eqnarray}
\frac{d^2(E\sigma)}{dE^2}&=&\pi R_B^2\,\delta(E-B) \nonumber \\
&\rightarrow&
\pi R_B^2\,\left[\frac{2\pi}{\hbar \omega}\frac{e^x}{(1+e^x)^2}\right],
\end{eqnarray}
where $x=(2\pi/\hbar\omega)(B-E)$ and $\theta$ is the Heaviside step function.
These functions (especially the second) have been extremely important 
in the development of the notion of a fusion barrier
distribution~\cite{rss,dasg-review}, where the 
fact that ${d^2(E\sigma)}/{dE^2}$ is strongly peaked near to the barrier
is a crucial point.
\begin{figure}[ht!]
\centering
\includegraphics[width=0.40\textwidth,angle=0,clip]{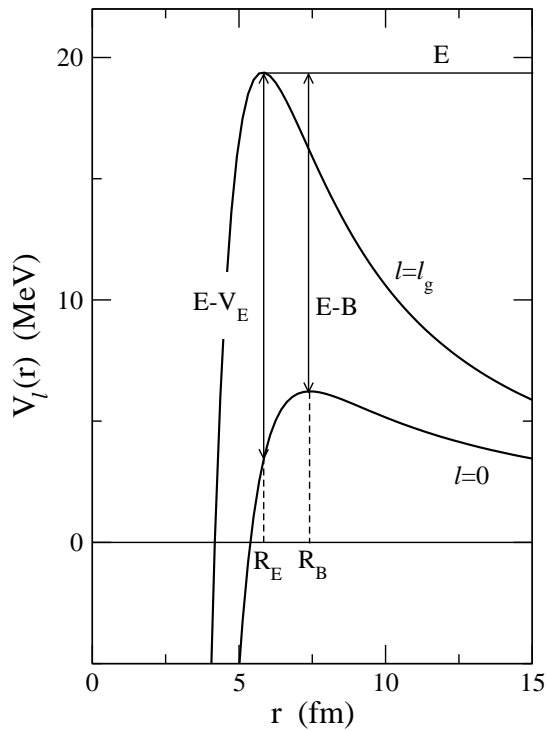}
\vspace*{0mm}
\caption{For an incident energy $E$ above the Coulomb barrier~$B$, 
the radius for the barrier with grazing angular momentum $l_g$
occurs at a separation $R_E<R_B$. The sum of the Coulomb and nuclear 
potentials at this point is $V_E<B$. Furthermore, the curvature $\hbar\omega_E$ 
of the barrier in the total potential is larger than its value for $l=0$.
The curves are calculated for an exponential
potential with $a=0.8$~fm and a depth
that yields $B=6.22$~MeV for the
$^{12}$C+$^{12}$C system. For $l=0$ this potential has $R_B=7.44$~fm and
$\hbar\omega_E=2.52$~MeV.
}
\label{fig:potential}
\end{figure}
\begin{figure}[ht!]
\centering
\includegraphics[width=0.40\textwidth,angle=0,clip]{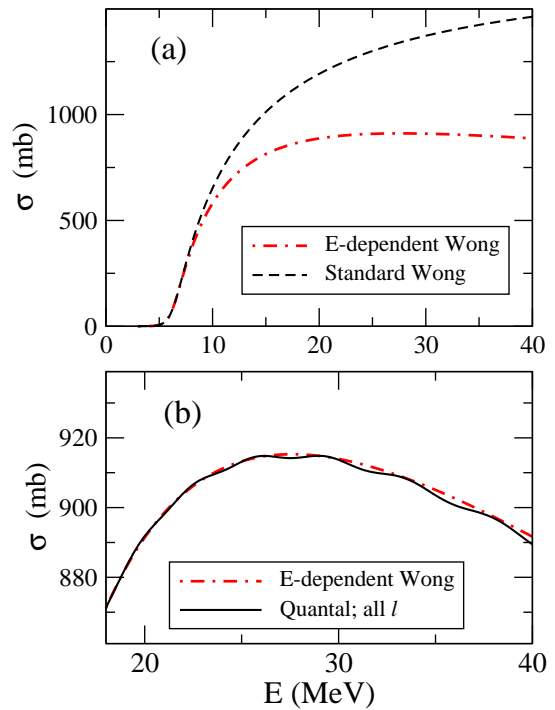}
\vspace*{0mm}
\caption{(Color online) 
  (a) Fusion cross sections for the $^{12}$C+$^{12}$C system obtained with 
the Wong formula with $l=0$ values of $[B,R_B,\hbar\omega]$ 
(dashed curve) and with their energy-dependent
values (dot-dashed curve). The Bose symmetry of the identical spin-zero system
is ignored, and both even and odd partial waves are summed up in the
cross sections. 
(b)~The energy-dependent Wong result agrees extremely well with a full quantal
calculation shown by the solid line.
However, the latter shows very weak oscillations ($\approx 1$~mb) 
coming from the partial-wave sum~(\ref{eq:sum}) even though all $l$ are summed. 
}
\label{fig:wong}
\end{figure}

It is important to notice that
higher above the barrier, one must question the approximations 
that lead to Eq.~(\ref{eq:wong0}). Figure~\ref{fig:potential}
shows that for higher angular momenta the barrier
in the full potential (Coulomb + nuclear + centrifugal) occurs at a separation
$R_E$ smaller than its $l=0$ value, $R_B$.
Now the grazing angular momentum 
is given by 
\begin{eqnarray}
\frac{l_g(l_g+1)\hbar^2}{2mR_E^2}=E-V_E,
\label{eq:graze}
\end{eqnarray}
where $V_E$ is the sum of the Coulomb and nuclear potentials at $R_E$. 
Of course the curvature changes too, not only because the position of the
barrier has shifted, but also because there is now a centrifugal contribution:
\begin{eqnarray}
  \hbar\omega_E=\hbar\left({-\frac{V''}{m}}\right)^{1/2}
  =\hbar\left({-\frac{V_C''+V_N''+V_l''}{m}}\right)^{1/2},
\label{eq:vpp}
\end{eqnarray}
where the second derivatives are evaluated at $r=R_E$. 
A~better approximation for the fusion cross section is now just
\begin{eqnarray}
\sigma=\frac{\hbar\omega_E}{2E}R_E^2\,
\ln\left(1+\exp\left[\frac{2\pi}{\hbar\omega_E}(E-V_E)\right]\right),
\label{eq:wong}
\end{eqnarray}
which is simply the standard Wong formula but with an energy-dependence 
$[B,R_B,\omega]\rightarrow [V_E,R_E,\omega_E]$ derived for the
grazing angular momentum $l_g$ at the energy in question. 
See Ref.~\cite{kabir-E-dep} for a similar extension to the classical
formula, Eq.~(\ref{eq:fus_cl}).

Of course we are still making an approximation here, because it is clear from 
Fig.~\ref{fig:potential} that the barrier height, position and curvature 
depend explicitly on $l$.
Of course if one does not assume the same values of the parameters for 
all~$l$, then the integral over $l$ leading to Eq.~(\ref{eq:wong})
cannot be performed analytically. But there is nothing to prevent
one choosing a different parameter set (for all~$l$) at each energy.
Clearly the set of values for $l_g$ is the best choice. One then still
obtains the compact expression~(\ref{eq:wong}) though we should stress that
one now needs to derive numerically the parameters $[V_E,R_E,\omega_E]$ (see
Appendix A).

We show in Fig.~\ref{fig:wong} some calculations for the system 
$^{12}$C+$^{12}$C that we shall concentrate on in this paper. 
For the moment, they ignore the Bose symmetry of this identical spin-zero 
system and sum over all even {\em and odd} partial waves.
The dashed line in Fig.~\ref{fig:wong}~(a) shows the Wong cross section 
with $[B,R_B,\omega]$ fixed at all energies to their values for $l=0$. 
That is  $[B,R_B,\omega]=[6.22~{\rm MeV},7.42~{\rm fm},2.52~{\rm MeV}]$,
generated by a potential with $a=0.8$~fm that gives this barrier height.
The results are seen to be very different from the dot-dashed line that uses the
energy-dependent values of these parameters. In Fig.~\ref{fig:wong}~(b)
we compare the latter results with a quantum mechanical calculation 
of the cross section. The scale of the vertical axis has been chosen to emphasise 
the high quality of the fit when the parameters are energy dependent. However,
it also shows that even with all partial waves, there are small oscillations in
the full quantal calculations. We shall derive an expression for these in the next
section.

\begin{figure}[ht!]
\centering
\includegraphics[width=0.40\textwidth,angle=0,clip]{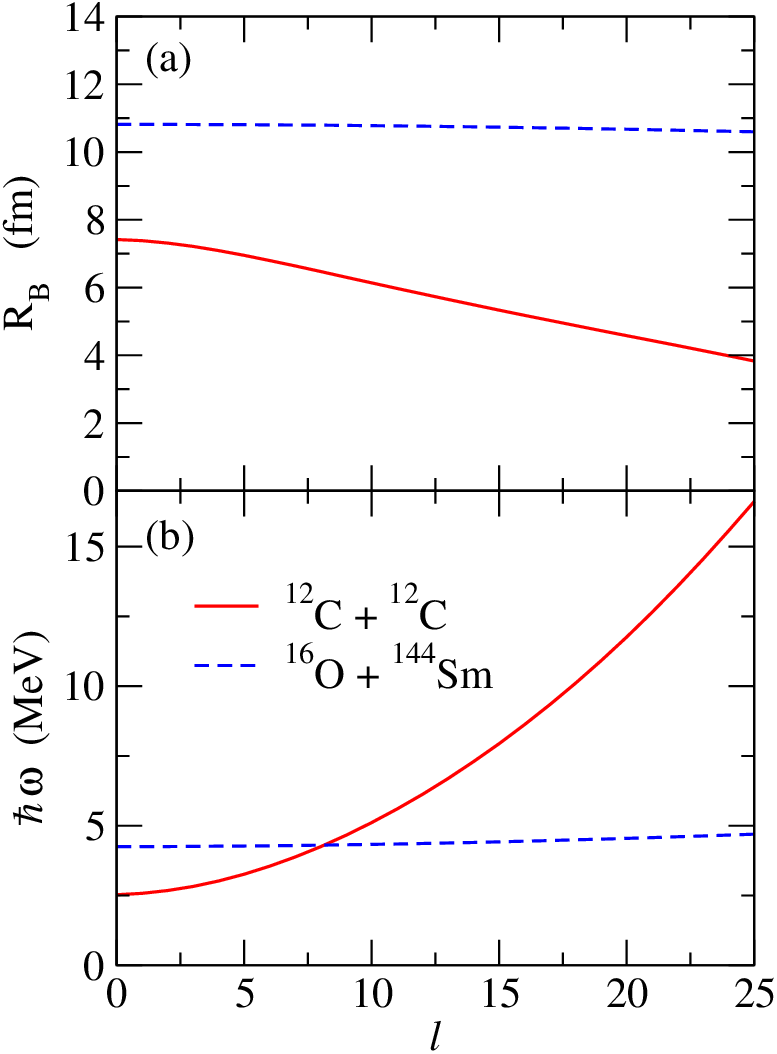}
\vspace*{0mm}
\caption{(Color online) The barrier position, $R_B$, (upper panel) 
and the barrier curvature, $\hbar\omega$, (lower panel) as a function 
of angular momentum, $l$, for the $^{12}$C+$^{12}$C system (solid line) 
and the $^{16}$O+$^{144}$Sm system (dashed line).} 
\label{fig:rb}
\end{figure}

The following comments are appropriate at this point:

\begin{itemize}
\item The approximation with parameters fixed
at their $l=0$ values is poor, especially for light systems. 
For these systems, the Coulomb interaction is relatively 
small and also the reduced 
mass $m$ is small, therefore the centrifugal potential may play a 
more important role than the Coulomb potential. For heavier systems, 
on the other hand, 
the Coulomb potential is strong and thus the barrier is rigid 
against a variation of angular momentum. In that situation, the 
conventional Wong formula is reasonable. Figure~\ref{fig:rb} shows 
the barrier position $R_B$ and the barrier curvature $\hbar\omega$ 
as a function of angular momentum $l$ for two different systems, 
$^{12}$C+$^{12}$C and $^{16}$O+$^{144}$Sm. One can see that the variation 
is marginal for the heavier system, $^{16}$O+$^{144}$Sm, while both 
$R_B$ and $\hbar\omega$ change considerably as a function of $l$ for the 
lighter system, $^{12}$C+$^{12}$C. 

\item 
However, note that experimental fusion cross sections (especially if they 
are not terribly precise) can frequently be fitted 
even for light heavy-ion systems 
with parameters that 
are simply chosen to do so, 
and do not necessarily bear much relation to
the physical potential. 
For example, we shall see in Sec.~III-D that the $^{12}$C+$^{12}$C
experimental fusion cross section can be reasonably well fitted with 
$[B,R_B,\hbar\omega]=[5.6,6.3,3.0]$ (see Fig.~\ref{fig:fits0} below). 

\item The $E$-dependent Wong formula works well because the parabolic 
approximation is good, since the energy $E$ coincides with the barrier height 
for $l=l_g$. Even though this approximation breaks down at energies 
below the barrier,  
only a small number of partial waves in the vicinity of $l_g$ contribute.
The cross section is relatively large above the barrier and the 
resulting discrepancies are small.

\item At sub-barrier energies, there is no $l_g$ as such,
and the best one can do is to take the parameters for $l$=0, though one is 
already below the barrier for this value of $l$. 
For higher $l$ values, the situation is worse, but 
taking an $l$-dependent value of the parameters 
does not improve matters, as the parabolic approximation
is in any case intrinsically poor below the Coulomb barrier 
\cite{HRD03,HT12}. 
Here the cross section is small, so the errors
are relatively more important. 
\end{itemize}

\section{Fusion oscillations}

\subsection{Fusion of two identical nuclei}

In most situations, the value of $\hbar\omega$ is irrelevant 
at energies above the Coulomb barrier,
simply determining the rate of fall-off for $E<B$. 
(Note that in that region the 
parabolic approximation itself will become inadequate at sufficiently 
low energies~\cite{HRD03,trotta}.)
However, (as first pointed out by Poff\'e {\em et al.}~\cite{poffe}, 
and more recently in Refs.~\cite{esbensen,wong-osc})
there is a rather unique situation where $\hbar\omega$ is important {\em above}
the barrier, principally for light symmetric 
spin-zero systems (though see below for other examples). 
Here, only even values of the grazing angular momentum are allowed and their
barriers may be sufficiently well
separated in energy for their successive addition to the cross section to give
rise to fusion oscillations.  

Let us now derive the earlier expression of Poff\'e {\em et al.}
~\cite{poffe} for these fusion oscillations. 
This is obtained by using the {\em exact} Poisson summation 
formula~\cite{poisson} 
\begin{eqnarray}
\sum_l f(l)&=&\sum_m (-1)^m\int_0^\infty f(\lambda)\,{\rm exp}(2\pi\,mi\lambda)\, d\lambda. \nonumber \\
\label{eq:poisson}
\end{eqnarray}
When applied to Eq.~(\ref{eq:hill}), 
the $m=0$ term gives the usual Wong result (\ref{eq:wong}). 
The terms $m=\pm 1$, however, give rise to an 
oscillatory contribution from the poles of the transmission function above
and below the real-$\lambda$ axis respectively. 
In the energy region $E-B\gg\hbar\omega/2\pi$, 
the nearest poles simply give (see Appendix B)
\begin{eqnarray}
E\sigma_{\rm osc}=2\pi R_E^2\hbar\omega_E\,{\rm exp}(-2\xi)\,
{\rm sin}(2\pi l_g),
\label{eq:poles}
\end{eqnarray}
where $l_g$ is the continuous, energy-dependent grazing angular 
momentum of Eq.~(\ref{eq:graze}),
and the quantity $\xi$ is given by
\begin{eqnarray}
\xi=\frac{\hbar\omega_E}{2l_g+1}\cdot\frac{\pi mR_E^2}{\hbar^2}\equiv\frac{\pi}{2}
\frac{\hbar\omega_E}
{\partial V_E/\partial l_g}.
\label{eq:xi}
\end{eqnarray}
The resultant oscillations are shown on top of the smooth part of 
the cross section in Fig.~\ref{fig:wong}~(b). They are of the order of
1 mb, and it is unlikely that any reasonable experiment would be able to observe these.
Terms from $|m|>1$ and from more distant poles merely introduce higher multiples of the 
exponent in the above equation and are clearly completely negligible. 

\begin{figure}[tb]
\vspace*{0mm} \centering
\includegraphics[width=0.4\textwidth,angle=0,clip]{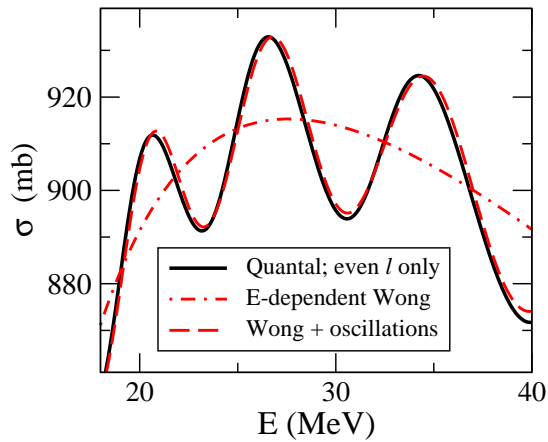}
\vspace*{0mm}
\caption{(Color online) 
The symmetrised cross section (even $l$ only) for $^{12}$C+$^{12}$C 
is compared with 
the cross section for all~$l$. One sees an order of magnitude difference
in the oscillatory terms (see text). 
It is unlikely that the oscillations shown in Fig.~2~(b) ($\approx$ 1 mb)
could be observed in a system without some asymmetry between 
odd and even $l$ values.
}
\label{fig:osc}
\end{figure}

However, for a system of two identical spin-zero bosons (for example, 
$^{12}$C+$^{12}$C)
the Bose symmetry forbids odd angular momenta, and the fusion cross section is
given by twice the sum over the even partial waves.  
This can be achieved by including the factor 
\begin{eqnarray}
1+[-1]^l\equiv 1+{\rm exp}(\pi il)
\end{eqnarray}
in Eq.~(\ref{eq:poisson}). The 1 gives the usual non-symmetrised result but the 
extra term now gives an oscillatory contribution (from $m=0$ and $m=-1$), 
\begin{eqnarray}
E\sigma_{\rm osc}=2\pi R_E^2\hbar\omega_E\,
{\rm exp}(-\xi)\,{\rm sin}(\pi l_g),
\label{eq:osc}
\end{eqnarray}
which is significantly larger than before, because the (negative) exponent 
has been reduced by a factor~2. This is demonstrated in Fig.~\ref{fig:osc}, 
where the dashed curve represents the 
sum of the smooth, energy-dependent Wong cross section and the 
above expression for the oscillatory part.  
The solid curve is the full quantal calculation, and it can be 
seen that the
above expressions give an excellent approximation to the exact result. 
The dot-dashed curve shows 
the smooth part of the Wong cross section.  
One sees an order-of-magnitude increase in the oscillations arising from 
symmetrisation.

The above result is easily generalised to two identical 
spin-1/2 fermions such
as  $^{13}$C+$^{13}$C. Because the oscillatory cross section for odd partial 
waves is clearly just minus that for the even ones, the total 
$\sigma_{\rm osc}$ is simply given
by the difference of the statistical weights for $S=0$ (which gives 
even $l$ with weight 1/4) 
and for $S=1$ (giving odd $l$ with weight 3/4). Thus we see that the 
oscillations 
will be reduced by a factor~2, and will have the opposite phase 
from the symmetric system.
Similarly, the oscillations for $^{14}$N +  $^{14}$N ($1\otimes 1$)
will be reduced by a factor of 3~\cite{poffe}. 
 
\subsection{Fusion between similar nuclei: the role of elastic transfer}

The other interesting problem of this type is that of a system 
such as $^{12}$C+$^{16}$O, which has also been discussed earlier by Kabir, 
Kermode and Rowley~\cite{kabir}.  
Here, one can again consider the oscillations from even $l$ and odd $l$
separately. Without any further effect, they will cancel out. However, 
the presence of the elastic $\alpha$-transfer channel will introduce a 
parity dependence into the 
problem. This is most easily seen by considering the total elastic scattering, 
where to the amplitude $f_{\rm el}(\theta)$ for direct elastic scattering 
of the $^{12}$C, we must add 
an exchange term $f_{\rm trans}(\pi-\theta)$, the amplitude for elastic 
transfer at the angle and energy in question. This yields a total amplitude
\begin{eqnarray}
f_{\rm total}(\theta)&\rightarrow& f_{\rm el}(\theta)+f_{\rm trans}(\pi-\theta),
\end{eqnarray}
and using $P_l({\rm cos}(\pi-\theta))=(-1)^lP_l({\rm cos}\,\theta)$ we obtain
\begin{eqnarray}
S_l^{\rm eff}=
S_l^{\rm \,el}+(-1)^lS_l^{\rm \,trans},
\end{eqnarray}
that is, different effective $S$-matrix elements for the odd and 
even partial waves. 
This effect can be simulated (as in Refs.~\cite{kabir,O70,OB75,VD86,LV87,CNV95}) 
by introducing a parity-dependent optical potential. 
However, to obtain a simple orientation of the effect of transfer
(and again an analytic expression for the oscillations) we 
note that the $S$-matrix for direct reactions is generally peaked around the grazing 
angular momentum and can be approximated by $\alpha\,\partial S^{\rm \,el}/\partial l$. 
For relatively small $\alpha$ (which is of course limited by unitarity), this  
gives an effective $S$-matrix
\begin{eqnarray}
S^{\rm \,eff}(l)=S^{\rm \,el}(l+[-1]^l\,\alpha).
\end{eqnarray}
That is, the transfer dynamics results in a small shift of the original elastic
$S$-matrix in {\em opposite directions} for odd and even $l$ values.
Clearly for a non-zero $\alpha$ the poles giving rise to Eq.~(\ref{eq:osc}) come into
play and indeed for $\alpha$=1/2, we obtain the same magnitude oscillations with a shift in
phase. The general result is that the trigonometric function in Eq.~(\ref{eq:osc}) is replaced
as follows
\begin{eqnarray}
{\rm sin}(\pi l_g)&\rightarrow& \frac{{\rm sin}(\pi [l_g+\alpha])-{\rm sin}
(\pi [l_g-\alpha])}{2} \nonumber \\
&=&{\rm cos}(\pi l_g){\rm sin}(\pi \alpha).
\label{eq:eltransfer}
\end{eqnarray}
Of course in this expression $\alpha$ may be energy dependent, and indeed if one uses a 
parity-dependent potential, this will naturally give rise to a shift that depends on 
the grazing $l$ and thus on $E$.
Ref.~\cite{kabir} demonstrated that it is possible to obtain a
reasonable fit to both the large-angle elastic scattering and the fusion oscillations
with the same parity-dependent potential.

\subsection{Heavier systems}

The presence or absence of measurable fusion oscillations arising from the
symmetrization for identical systems depends mainly on the quantity~$\xi$ in 
Eq.~(\ref{eq:xi})
because it appears in the exponent in Eq.~(\ref{eq:osc}). Using an
exponential nuclear potential, Eq.~(\ref{eq:vpp}) yields exactly
\begin{eqnarray}
\hbar\omega_E
=\hbar\left({\frac{2E-V_E-\frac{a}{R_E}\varepsilon}{a\,m\,R_E}}\right)^{1/2}, 
\end{eqnarray}
with $\varepsilon\equiv 4(E-V_E)+V_{CE}$, where $V_{CE}$ is the Coulomb potential at $R_E$.
Close to the s-wave barrier $B$, the nuclear potential is relatively small, and there
$\varepsilon\approx 4E-3V_E$. However, at higher energies and higher grazing $l$ values, the
barrier is pushed to lower radii and the nuclear potential gives the dominant contribution.
Here $V_{CE}$ can be neglected so that $\varepsilon\approx 4(E-V_E)$ yielding
\begin{eqnarray}
\xi\approx\frac{\pi}{2}\sqrt{\frac{R_E}{2a}}
\left(\frac{2E-V_E}{E-V_E}-\frac{4a}{R_E}\right)^{1/2} 
\label{eq:xi-heavy}
\end{eqnarray}
or in terms of $B$
\begin{eqnarray}
\xi\approx\frac{\pi}{2}\sqrt{\frac{R_E}{2a}}
\left(1-\frac{4a}{R_E}+\left[1-\frac{2a}{R_E}\right]\frac{E}{E-B}\right)^{1/2}. 
\label{eq:xi-heavy-B}
\end{eqnarray}
Asymptotically $(E\gg B)$ this reduces to 
\begin{eqnarray}
\xi\approx\frac{\pi}{2}\sqrt{\frac{R_E}{a}}\left(1-\frac{3a}{R_E}\right)^{1/2},
\label{eq:xi-asymp}
\end{eqnarray}
and this gives a good idea of the mass dependence of the magnitude of the 
resulting oscillations.
For heavy systems, $R_E$ is larger, and so is $\xi$. Thus the 
oscillations will become more difficult to observe experimentally in heavier systems. 

\begin{figure}[ht!]
\begin{center}
\includegraphics[width=0.4\textwidth,angle=0,clip]{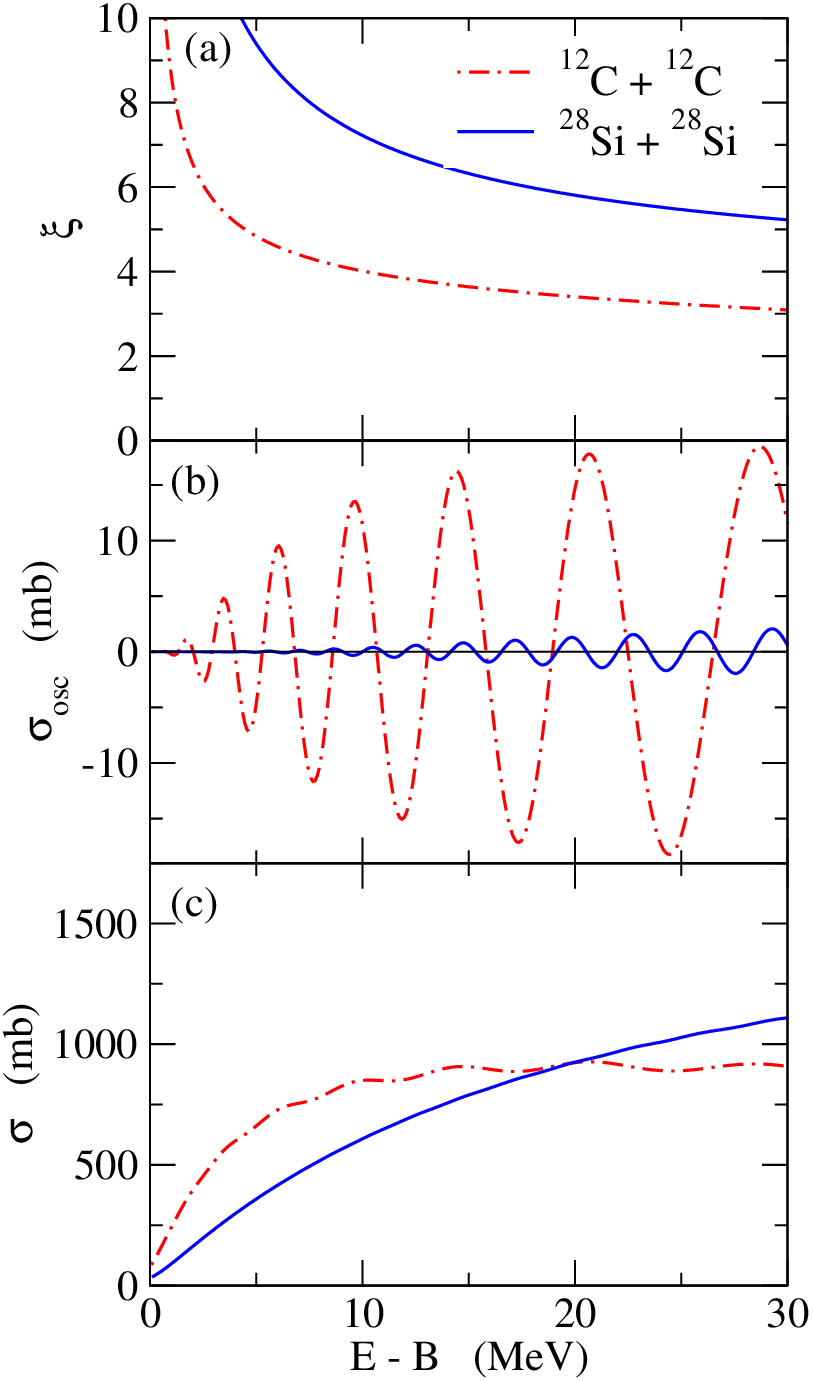}
\caption{(Color online) (a) The $\xi$ parameter 
defined by Eq.~(\ref{eq:xi}) 
in the oscillatory part of fusion cross 
sections. The dot-dashed and the solid lines show the $\xi$ parameters 
for the $^{12}$C+$^{12}$C system 
the $^{28}$Si+$^{28}$Si system, respectively, as a function of the energy 
measured from the barrier height $B$ of each system. 
Both lines are obtained with an exponential potential with the diffuseness 
parameter of $a$=0.8~fm. 
(b) The oscillatory contribution of fusion cross sections 
given by Eq.~(\ref{eq:osc}). 
(c) The total fusion cross sections for the two systems 
given as a sum of the smooth part and the oscillatory 
part. 
}
\label{fig:c-si}
\end{center}
\end{figure}

The top panel of Fig.~\ref{fig:c-si} compares 
the value of $\xi$ for the $^{12}$C+$^{12}$C system 
(dot-dashed line) with that for $^{28}$Si+$^{28}$Si 
(solid line). For the former reaction, we have used the same potential as in 
Fig.~\ref{fig:wong}, while we have used an exponential potential with 
$[B,a]=[28.8 {\rm ~MeV},0.8 {\rm ~fm}]$ for the latter system. 
Even though the $\xi$ parameter for 
$^{28}$Si+$^{28}$Si is larger than that for the 
$^{12}$C+$^{12}$C system by a factor of only about~2, 
its effect on the amplitude of the fusion oscillations is dramatic, as seen 
in the middle panel of Fig.~\ref{fig:c-si}. The oscillations shown here
are obtained from Eq.~(\ref{eq:osc}) but they can also be simply obtained from 
the quantum mechanical sum over~$l$: because the oscillatons arising from
symmetrization are exponentially larger that those with no symmetry, they
are essentially given by the difference between these two cross sections. This 
reduces to
\begin{eqnarray}
\sigma_{\rm osc}=\frac{\pi\hbar^2}{2mE}\left(\,\sum_{l~{\rm even}} (2l+1)T_l - 
\sum_{l~{\rm odd}} (2l+1)T_l\right),
\label{eq:sum-osc}
\end{eqnarray}
where we note that this result is true only if the sum over all partial 
waves is sufficiently smooth. 

Of course experimentally it is not possible to separate the oscillations
out of the total cross section, and although they stand out in 
the $^{12}$C+$^{12}$C system (see bottom panel of  Fig.~\ref{fig:c-si}), 
they are not apparent in the total cross section for $^{28}$Si+$^{28}$Si. 
For this reason the relevant experimental data are sometimes represented in the 
form of $d(E\sigma)/dE$~\cite{esbensen}. This is a useful representation since the derivative
of the smooth part of  $E\sigma$ is essentially constant. We shall look at this again below.

\begin{figure}[ht!]
\begin{center}
\includegraphics[width=0.4\textwidth,angle=0,clip]{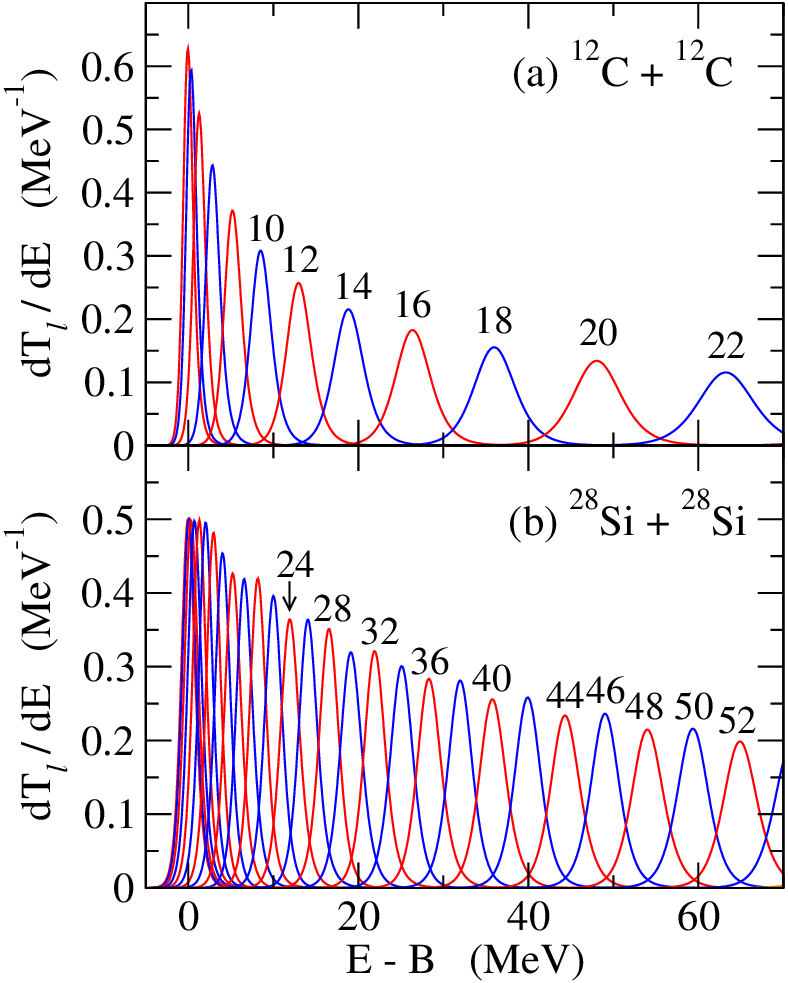}
\caption{(Color online) 
The first energy derivative of the penetrability for each angular 
momentum $l$, $dT_l/dE$, for the $^{12}$C+$^{12}$C system (upper 
panel) and for the $^{28}$Si+$^{28}$Si system (lower panel). 
The numbers above the peaks are the corresponding values of $l$. 
}
\label{fig:dpde}
\end{center}
\end{figure}

In the middle panel of Fig.~\ref{fig:c-si}, we have seen that 
the oscillations for the $^{28}$Si+$^{28}$Si system are largely damped out
at energies less than around 10~MeV above the barrier, while they 
start much earlier, and are much stronger, for the $^{12}$C+$^{12}$C system.  
In order to understand this, Fig.~\ref{fig:dpde} compares 
the derivative of the penetrability for each angular momentum
with respect to the energy, 
that is, $dT_l/dE$, for the $^{12}$C+$^{12}$C system (upper panel) 
with that for the $^{28}$Si+$^{28}$Si system (lower panel).  
For the latter system, the individual peaks are narrower (due to smaller values of  
$\hbar\omega$), but they are pushed much closer together, making them less well 
resolved from the peaks for adjacent values of $l$. In order to obtain the 
same resolution that one has for $l=12$ in the $^{12}$C+$^{12}$C system
(that is, the same degree of overlap between adjacent peaks), 
one has to go to an $l$ value of around 36 or bigger 
in the $^{28}$Si+$^{28}$Si system. This corresponds to a much larger energy 
compared with the former system, and indeed the limiting angular momentum
for fusion for $^{28}$Si+$^{28}$Si appears to be 
$l_{\rm max}=38$, according to Vineyard {\it et al.}~\cite{vineyard}.

\begin{figure}[ht!]
\begin{center}
\includegraphics[width=0.4\textwidth,angle=0,clip]{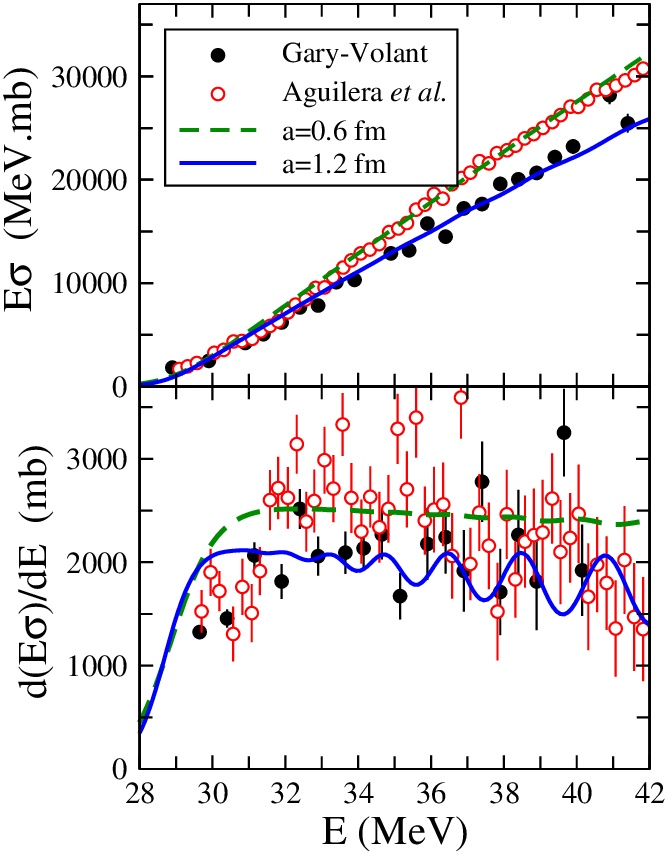}
\caption{(Color online) 
Two different sets of measurements~\cite{aguilera,gary-volant} 
for fusion in the $^{28}$Si+$^{28}$Si system are 
shown as  $E\sigma$ in the upper panel and as $d(E\sigma)/dE$ in the lower one.
The data are compared with uncoupled calculations with two different values of
surface diffuseness in the nuclear potential. See text.
}
\label{fig:28si28sidata}
\end{center}
\end{figure}

Several other data sets for the $^{28}$Si+$^{28}$Si system
exist~\cite{nagashima,gary-volant,aguilera} and indeed one of the aims
of Aguilera {\it et al.}~\cite{aguilera} was to search for oscillations in this system.
Unfortunately they were unable to conclude the existence of structures within the 3\% 
statistical error in their experiment. Their data are represented in the form $E\sigma$ in the upper
panel of Fig.~\ref{fig:28si28sidata} (open circles), along with data from 
Ref.~\cite{gary-volant} (solid circles). It can be seen that the normalizations of the
two data sets disagree somewhat, presumably simply due to systematic differences in the experiments.
Also shown in the figure are two uncoupled optical-model calculations with 
$a$=0.6~fm and 1.2~fm (both with $B$=28.7~MeV). 
The slope of the former fits that of the Aguilera data but the larger diffuseness 
is required to fit the slope of the data of Ref.~\cite{gary-volant}.
In the lower panel of Fig.~\ref{fig:28si28sidata}
the two data sets and calculations are again shown but in terms of $d(E\sigma)/dE$. 
Neither data set has the precision to prove the existence of oscillations, but it is 
interesting to note the significant difference in the magnitude of the oscillations
produced by the different values of the potential diffuseness $a$ (see Eq.~(\ref{eq:xi-asymp})). 
Better data are clearly required to resolve this question, and to establish the
existence or absence of any oscillations due to symmetrization.

Very recently Stefanini {\it et al.}~\cite{Stefanini14} have made 
more detailed measurements for this system with a very small energy step 
(around 0.25~MeV in the center-of-mass system) and with better statistics
(better than 1\%). Their experimental data, when displayed as $d(E\sigma)/dE$,
indicate distinct structures at energies only around 5~MeV above the barrier.
They are much more pronounced than those shown for the uncoupled calculation
with  $a$=1.2~fm in Fig.~\ref{fig:28si28sidata}. 
This observation is incompatible with the mechanism discussed here, and 
the oscillations in the new experimental data must have a different origin. 
Their presence appears to be related to strong couplings to the low-lying collective modes
that exist in the $^{28}$Si nucleus, but also seems to depend on the nature
of the absorptive potential used in the calculations~\cite{esbensen,Stefanini14}. 

In order to look for the effects of channel couplings, the `experimental barrier
distribution'~\cite{rss,dasg-review}  $D(E)=d^2(E\sigma)/dE^2$ is frequently used, and 
in the present context one should remember that the function $d(E\sigma)/dE$ 
can also have structures close to the unperturbed Coulomb barrier 
due to couplings. The width of a typical barrier distribution is proportional
to $Z_1Z_2$ and will lead to obvious structures in $d(E\sigma)/dE$ for heavier
systems with strong collective modes. However, for heavier systems,
the `symmetrization' oscillations will not be measurable. The system 
$^{28}$Si+$^{28}$Si may be a special intermediate system where both types
of structure are present simultaneously.

\section{Comparison with experimental data on lighter systems}

\subsection{$^{12}$C+$^{12}$C fusion reaction} 

Let us now analyze the actual experimental data for fusion 
of carbon isotopes and discuss the observed fusion oscillations. 
We first discuss the fusion of the $^{12}$C+$^{12}$C system using a 
single-channel optical model. 

We have already shown in Fig.~\ref{fig:rb}
the energy-dependence of the parameters $R_E$ and $\hbar\omega_E$ 
entering both the smooth part of the Wong cross section
and its oscillatory term
for the potential of Fig.~\ref{fig:potential}
that gives a good fit to the experimental $^{12}$C +  $^{12}$C
fusion cross section, including 
its oscillations. (Some of these energy variations have also been discussed 
in Ref.~\cite{kabir-E-dep},
where they are expressed in terms of universal functions.)
The function $dT_l/dE$ shown in the upper panel of Fig.~\ref{fig:dpde}
is for an optical-model
calculation with the same potential and various $l$ values.
Its width
is $0.56 \,\hbar\omega_E$~\cite{rss}, 
and this compares well with the values of $\hbar\omega_E$ from
Fig.~\ref{fig:rb}, taken at the appropriate peak
energies (where the $l$ values in question are grazing).
The variation of all of these Wong 
parameters is very significant over the energy range in question,
and it is important to remember 
that their $l=0$ values do not even fit the average data when
inserted into Eq.~(\ref{eq:wong}).

\begin{figure}[ht!]
\vspace*{0mm} \centering
\includegraphics[width=0.3\textwidth,angle=0,clip]{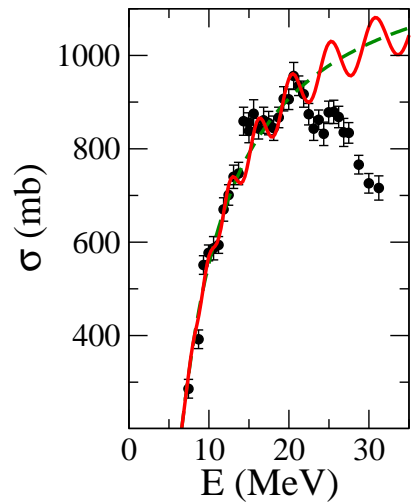}
\vspace*{0mm}
\caption{
(Color online)
  Parametrised $^{12}$C+$^{12}$C cross section~\cite{poffe,HT12} with
  $[B,R_B,\hbar\omega]=[5.6,6.3,3.0]$. The smooth-only cross section
  is shown by the dashed curve.
The experimental data are taken from Ref.~\cite{K79}. 
}
\label{fig:fits0}
\end{figure}

\begin{figure}[tb]
\vspace*{0mm} \centering
\includegraphics[width=0.47\textwidth,angle=0,clip]{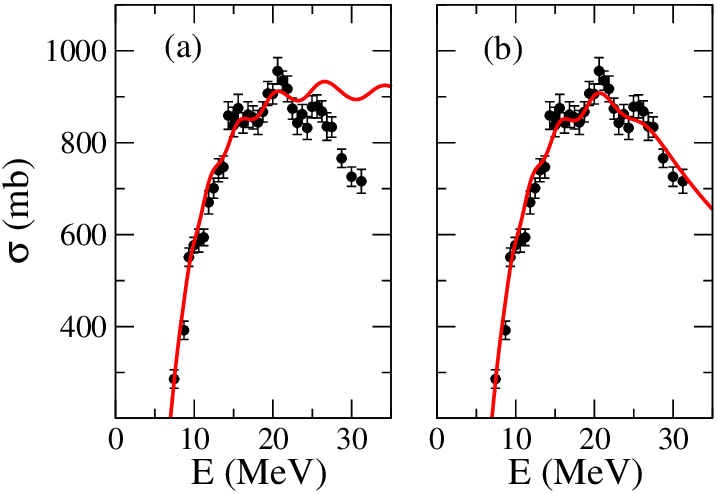}
\vspace*{0mm}
\caption{(Color online)
  (a) Potential-model fit to the same data as in Fig.~\ref{fig:fits0} 
  with $[B,a]=[6.22,0.8]$ (even $l$ only).
  (b)~Same as (a) but with the transmission of $l=14$ reduced by a
  factor~2 and all higher waves absent. 
}
\label{fig:fits}
\end{figure}

In Fig.~\ref{fig:fits0}, the data are reasonably well fitted~\cite{poffe,HT12} 
(at least up to around 20~MeV) with an energy independent
set of $[B,R,\hbar\omega]=[5.6,6.3,3.0]$
but these parameters do not correspond to those coming
from the potential-model fit shown in Fig.~\ref{fig:fits}~(a).
(Though the value of the oscillatory term
is similar to the potential-model value in the region where the oscillations are important, 
because the expression~(\ref{eq:osc}) was used for it.)
For instance, Eq.~(\ref{eq:exp-B}) indicates that this parameter set
corresponds to an exponential potential with the diffuseness parameter
of $a=2.0$~fm, that is unphysically large.
The Wong cross section can, therefore, be considered only
as providing a simple parametrisation
of the fusion data, although as pointed out earlier this
is extremely useful if one remains close
to the barrier. 

This point is well demonstrated by the failure to fit the cross
section at energies 
above $E\approx 25$~MeV, and Fig.~\ref{fig:fits}~(b) shows that a
likely explanation of this is the 
failure of higher partial waves to fuse. This fit is obtained with
the transmission $T_{l=14}$ reduced
by a factor of 2, and all higher partial waves completely
removed from the cross section. 

Such an assumption is not unreasonable, because in this region of
the compound nucleus spin $I$, the excitation energy of the $^{24}$Mg
formed in the $^{12}$C+$^{12}$C reaction at the barrier height 
corresponding to an angular
momentum $l=I$, is not sufficient for s-wave particle emission. That is,
\begin{eqnarray}
Q+E_{\rm \,barrier}(l) \equiv E^*_{\rm \,barrier}(I=l)<E_{\rm \,yrast}(I)+S_x,  
\label{eq:yrast}
\end{eqnarray}
where $Q$ is the reaction Q value (13.93~MeV), $E_{\rm yrast}$ is the yrast energy of
$^{24}$Mg~\footnote{The ground-state rotational band of $^{24}$Mg is well fitted up to the highest
identified spin of Ref.~\cite{wiedenhover} ($I^{\pi}$=$10^+$ at 19.2~MeV) by the Harris parametrization~\cite{harris} 
with ${\cal J}_0$=2.22 $\hbar^2.{\rm MeV}^{-1}$ and ${\cal J}_1$=0.115 $\hbar^4.{\rm MeV}^{-3}$. This gives the energies
$E_{12^+}$=25.80~MeV,  $E_{14^+}$=33.05~MeV, 
$E_{16^+}$=40.85~MeV\dots, and so forth.},
and $S_x$ is the appropriate particle separation energy (where $x\in$ neutron, $S_n$=16.53~MeV; 
proton, $S_p$=11.69~MeV; or alpha particle, $S_\alpha$=9.31~MeV).
But, for $l \ge 12$ we find $E^*_{\rm barrier}(I=l) > E_{\rm yrast}(I-1)+S_{\alpha}$, 
permitting the emission of
an $L=1$ alpha particle. Similarly for particle angular momentum $L=2$, proton emission also becomes possible. 
However, such emissions are inhibited by the penetration of the corresponding centrifugal barriers as well
as the relevant Coulomb barriers, and thus competition with fission will become important
in the spin range above $I$=12.

\begin{figure}[tb]
\vspace*{0mm} \centering
\includegraphics[width=0.47\textwidth,angle=0,clip]{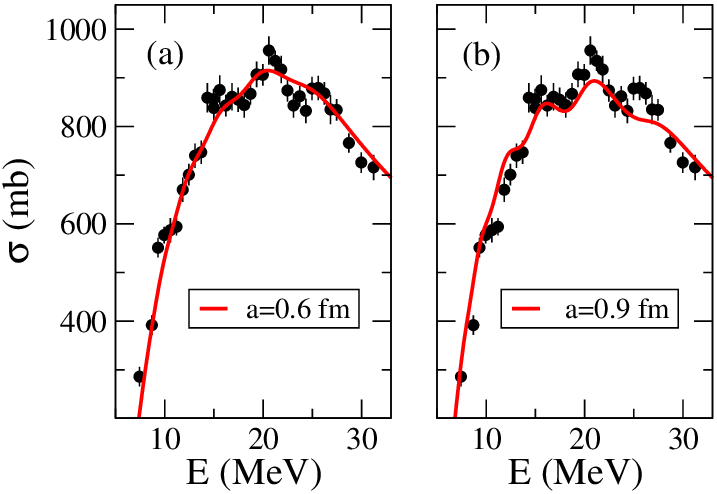}
\vspace*{0mm}
\caption{(Color online)
Fits to the $^{12}$C+$^{12}$C cross section with different potential 
parameters. The surface diffuseness $a$ is constrained both by the
magnitude of the smooth cross section and by the presence of oscillations. 
The left panel (a) shows results with
$a=0.6$~fm and $B=6.54$~MeV, while the right panel (b)
has $a=0.9$~fm and $B=6.06$~MeV.  
An intermediate value of $a=0.8$~fm appears best (see Fig.~\ref{fig:fits}).
}
\label{fig:fits2}
\end{figure}

In Fig.~\ref{fig:fits2} we show
potential model fits to the data with different values of the
surface diffuseness $a$.
While a good fit was obtained with $a=0.8$~fm as shown
in Fig.~\ref{fig:fits}~(b),
Figs.~\ref{fig:fits2}~(a) and (b) show the best fits
with slightly smaller and larger values $a=0.6$~fm and $a=0.9$~fm, 
respectively.
In each case, the value of $B$ is adjusted slightly to obtain the best fit
and a clear pattern emerges.
If one decreases $a$ this increases $R_B$
(Eq.~(\ref{eq:rb})) which in turn increases the 
higher-$E$ cross section.
This can be compensated by a slight increase in the barrier height which 
correspondingly decreases the radius again. Thus there is a
little `play' in the value of the potential
parameters but one cannot
stray too far from the `best' values without destroying the fit either at 
higher or lower energies. Furthermore, the presence of the
oscillations provides an additional strong constraint.
For the smaller $a$ value of $0.6$~fm,
the average cross section is fitted very well, but the 
magnitude of the oscillations is strongly damped.
For the larger $a$ value of $0.9$~fm, the 
magnitude of the oscillations is increased
(possibly improved) but the smooth part of the 
cross section starts to deviate. Thus we see that
the data provide strong physical constraints 
on the physical properties of the nucleus-nucleus
potential that are lost when the parameters
$[B,R_B,\hbar\omega]$ are regarded as independent variables.
Thus while the elegance and 
simplicity of the Wong cross section should be recognised,
so should its limitations.

\begin{figure}[tb]
\vspace*{0mm} \centering
\includegraphics[width=0.47\textwidth,angle=0,clip]{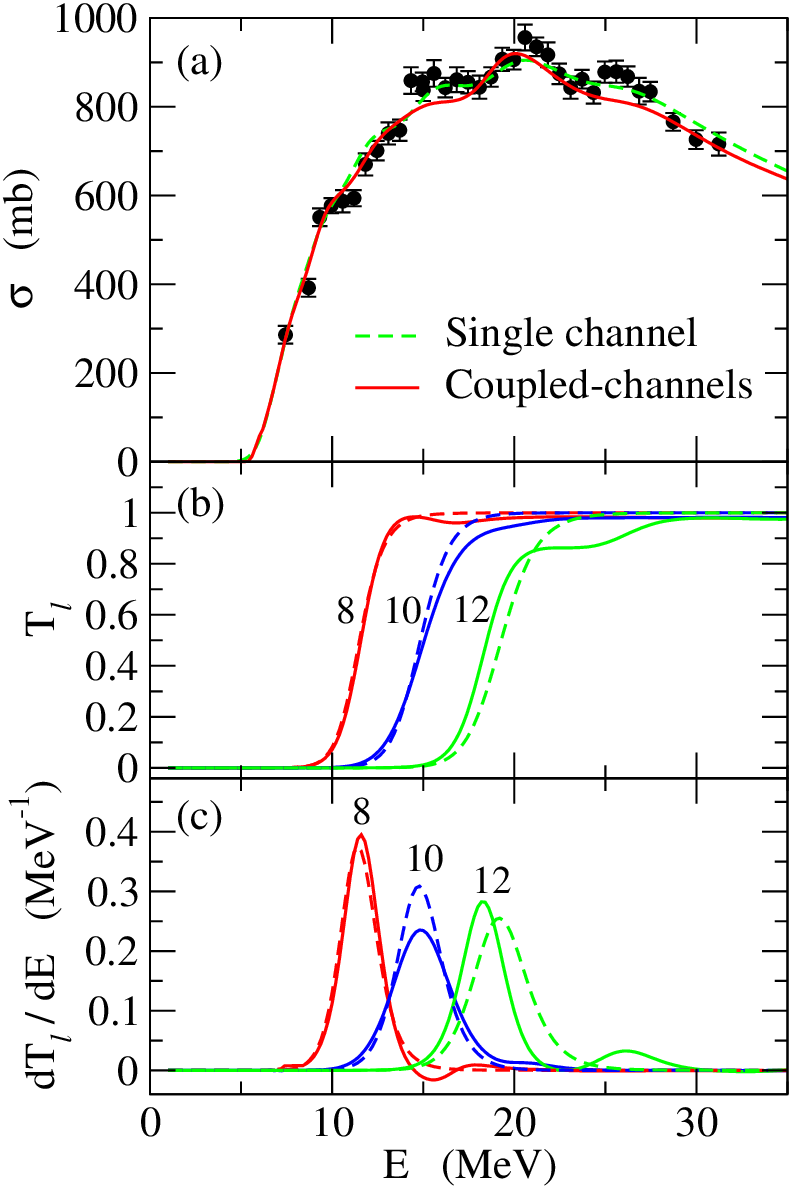}
\vspace*{0mm}
\caption{(Color online)
Effect of channel coupling on the fusion excitation function for  
the $^{12}$C+$^{12}$C system (top panel). 
The dashed line is the same 
as the solid line in Fig.~\ref{fig:fits} (b)~obtained with the potential 
model. The solid line, on the other hand, is obtained by including the 
rotational coupling to the first 2$^+$ state both in the projectile and 
the target nuclei. The depth parameter of nuclear potential is slightly 
adjusted in order to remove the trivial barrier renormalization due to 
the couplings. 
The penetrability $T_l$ and its first derivative $dT_l/dE$ 
are shown in the middle and the bottom panels, respectively, for 
$l=8,10$ and 12. 
}
\label{fig:12c12c-cc}
\end{figure}

So far we have discussed the fusion oscillations
for the $^{12}$C+$^{12}$C system 
based on the single-channel 
potential\-model calculations. In order to discuss the role of channel 
couplings, the top panel of Fig.~\ref{fig:12c12c-cc} shows 
results of a coupled-channels calculation 
for the $^{12}$C+$^{12}$C system. To this end, we include the rotational 
coupling to the first 2$^+$ state both in the projectile 
and the target nuclei with the deformation parameter 
of $\beta$=-0.40~\cite{yasue} (with a radius parameter
$r_0$=1.06~fm). In order to remove the potential renormalization due to the
coupling 
\cite{HT12,THAB94,HTDHL97}, 
we slightly adjust the depth parameter of 
the nuclear potential. We use a modified version of {\tt CCFULL} \cite{ccfull} 
to solve 
the coupled-channels equations. 
The solid and the dashed lines in Fig.~\ref{fig:12c12c-cc} denote the 
results of the coupled-channels and the single-channel calculations, 
respectively. One can see that the main feature of the fusion oscillations, 
including the peak energies and the phase of the oscillations, are not 
affected much by the channel-coupling effects, and thus our discussions 
based on the simple potential model calculations remain valid. 

This conclusion is due to the fact that
the barrier distributions for non-zero $l$ 
\cite{esbensen,rss,HT12} still show almost a single-peaked structure, 
as shown in Figs.~\ref{fig:12c12c-cc} (b) and~(c). This originates for 
the following two reasons. Firstly the excitation energy of the 2$^+_1$ 
state in $^{12}$C is relatively large ($E_2$=4.44~MeV) and thus the 
adiabatic approximation is good, leading to the adiabatic-barrier 
renormalization~\cite{HT12,THAB94,HTDHL97}. 
Secondly, with rotational coupling to an oblate nucleus,
the lowest barrier, which is relevant to 
the adiabatic-barrier renormalization, carries most of the weight, 
as in the vibrational case. If $^{12}$C had a prolate deformation, the 
lowest barrier would have the smallest weight in the barrier distribution, and 
the fusion oscillations would be much more affected by the couplings. 
In that case, the oscillations are significantly damped, 
and thus the data essentially determine 
the sign of the deformation of $^{12}$C. 

\subsection{$^{12}$C+$^{13}$C fusion reaction} 

\begin{figure}[tb]
\vspace*{0mm} \centering
\includegraphics[width=0.47\textwidth,angle=0,clip]{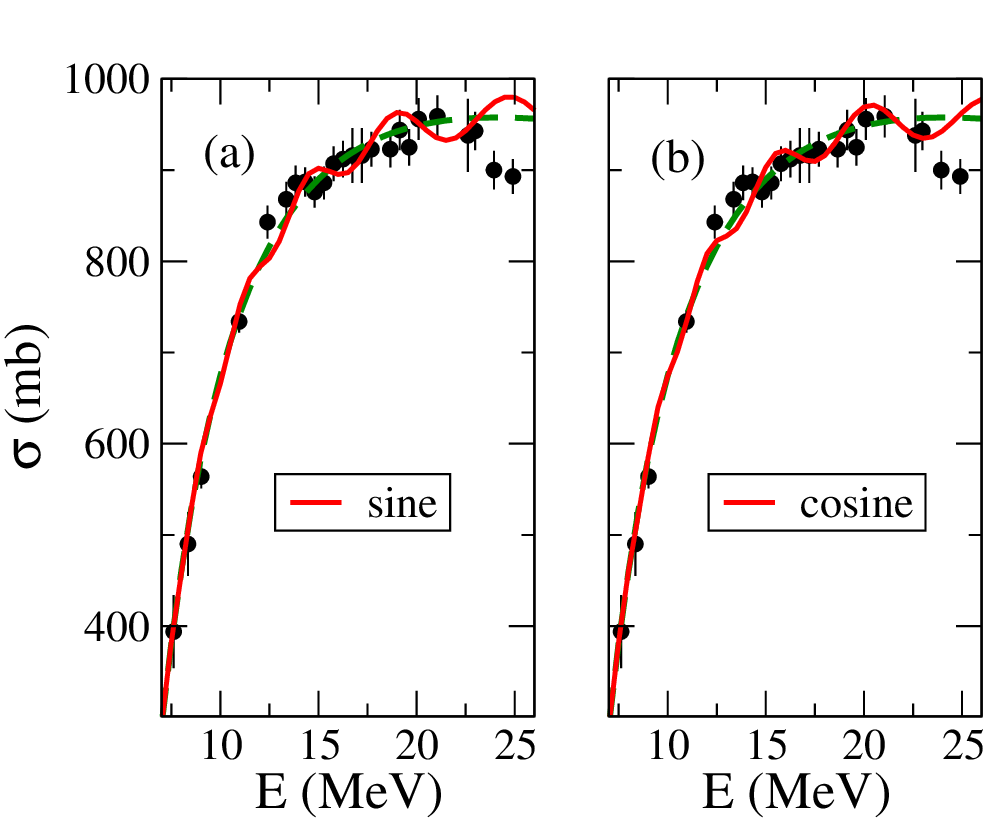}
\vspace*{0mm}
\caption{(Color online)
(a) Comparison between the experimental 
  fusion excitation function 
  and theoretical curves for the $^{12}$C+$^{13}$C system.
  The dashed line is obtained 
with the Wong formula, while the solid line shows 
a fit to the data using Eq.~(\ref{eq:osc}).
The experimental data are taken from Ref.~\cite{K79}. 
(b)~Same as~(a) but with Eq.~(\ref{eq:eltransfer}) and $\alpha$=$-$0.5. }
\label{fig:12c13c}
\end{figure}

Let us next discuss the $^{12}$C+$^{13}$C systems, for which the effect
of elastic neutron transfer is expected to play an important role. 
Based on the discussion in the previous subsection, we 
shall use the potential model for our discussions. 
Fig.~\ref{fig:12c13c} (a) shows the fusion cross section for this system 
and its fit with the Wong formula (dashed line). 
Although the oscillation is less prominent than for 
$^{12}$C+$^{12}$C, the fusion cross section still shows significant
oscillations and the Wong formula does not account well for them. 
The solid line in Fig.~\ref{fig:12c13c} (a) shows
a fit with Eq.~(\ref{eq:osc}). 
Although the magnitude of the oscillatory structure is well reproduced,  
the oscillation is out of phase with the experimental data. 
Fig.~\ref{fig:12c13c}~(b) shows a fit with Eq.~(\ref{eq:eltransfer}) 
with $\alpha$=$-$0.5. It is apparent that this fit is better than the 
other two, and thus the experimental data appear to determine the relative 
phase of the transfer and elastic $S$-matrix elements.  

\begin{figure}[tb]
\vspace*{0mm} \centering
\includegraphics[width=0.47\textwidth,angle=0,clip]{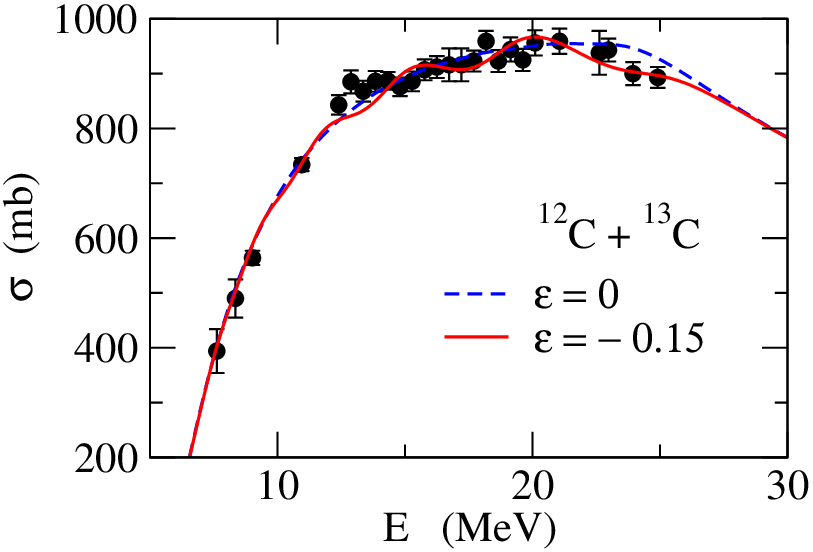}
\vspace*{0mm}
\caption{(Color online)
Fusion excitation function for $^{12}$C+$^{13}$C 
obtained with a parity-dependent potential. 
The parity dependence is introduced by replacing 
$V_0$ with $V_0(1+(-1)^l\epsilon)$ in the exponential potential 
$V_N(r)=V_0e^{-r/a}$, where $V_0$ takes a negative value. 
The solid and the dashed lines are obtained with $\epsilon=-0.15$ and 0, 
respectively. The angular momentum sum is truncated at $l=14$ as in 
Fig.~\ref{fig:fits} (b). }
\label{fig:12c13c-paritydep}
\end{figure}

Naturally, the sign of $\alpha$ is related to the sign of a parity-dependent 
part of optical potential. In fact, 
if one introduces a parity-dependent barrier height, $V_E\pm(-1)^l\Delta V$, 
it is easy to show that 
the grazing angular momentum $l_g$ is given 
to first order in $\Delta V$ by 
\begin{equation}
l_g=l_g^{(0)}
\mp (-1)^l\frac{\Delta V_E}{2}\,\sqrt{\frac{2m R_E^2}{(E-V_E)\hbar^2}},
\end{equation}
where $l_g^{(0)}$ is the grazing angular momentum for $\Delta V$=0.
This implies that $\alpha$ is given by
\begin{equation}
\alpha=
\mp (-1)^l\frac{\Delta V_E}{2}\,\sqrt{\frac{2m R_E^2}{(E-V_E)\hbar^2}}, 
\label{alpha}
\end{equation}
indicating that $\alpha$ is positive (negative) if 
the barrier
height for even-partial waves is lower (higher) than that for odd-partial 
waves. The negative sign of $\alpha$ for the $^{12}$C+$^{13}$C system suggests 
that the barrier for this system is lower for odd-partial waves.  
This is consistent with earlier findings in Refs.~\cite{O70,VD86}. 
In Fig.~\ref{fig:12c13c-paritydep}, we show fusion cross sections 
obtained with a parity-dependent potential. 
The parity dependence is introduced by using an energy-dependent 
depth parameter, $V_0(1+(-1)^l\epsilon)$ in the exponential potential 
(here, $V_0$ is defined as taking a negative value). The solid line in the 
figure indicates that the experimental data are well accounted for 
with $\epsilon$=$-$0.5, which indeed implies higher barriers for 
even partial waves. 

In contrast to $^{12}$C+$^{13}$C, the quantity
$\alpha$ has been found to be positive 
for the $^{12}$C+$^{16}$O system~\cite{kabir}, and we 
note that Baye~\cite{B86} has proposed a simple rule for the sign of the
parity-dependent potential 
in terms of the mass number $A_{\rm core}$ of the identical `cores' of the colliding nuclei 
and the parities $\pi_i$ of the valence orbitals of the transferred nucleons. 
His rule is based upon a microscopic resonating group method (RGM) within a two-center
harmonic-oscillator shell model. 
Essentially the barrier for even partial waves is 
deemed to be higher (or lower) than that for odd partial waves if the quantity
\begin{eqnarray}
-(-1)^{A_{\rm core}} \prod_{\rm i:valence} \pi_i
\label{eq:baye}
\end{eqnarray}
is positive (or negative). In both of our cases  $A_{\rm core}$ is even. 
For $^{12}$C+$^{16}$O the number of valence
particles is also even and this expression is negative. For the $^{12}$C+$^{13}$C reaction
we have a single, odd-parity valence neutron ($p_{1/2}$), and the expression is positive.
The sign of $\alpha$ determined from the fusion oscillations 
is consistent with these results for both of our systems. 

Note that the sign of the parity-dependent term, obtained from a fit 
to the angular distribution of elastic scattering, 
shows some ambiguity for several systems \cite{KRS83}. 
Fusion oscillations may offer a direct and perhaps better 
way to determine the sign. 
We note, however, that even though the negative value of $\alpha$
fits well the fusion oscillations and is consistent with the Baye's rule, 
a positive value appears more consistent with the data
for energies around 13~MeV (see Fig. \ref{fig:12c13c-paritydep}).
It would be interesting to re-measure fusion cross sections in this
energy region with higher precision in order to confirm whether
there exists a shift in phase of the oscillations as a
function of energy.

\section{Summary}

The Wong formula 
has been widely used to estimate 
fusion cross sections for a given single-channel potential as well as 
to discuss the parameters which govern fusion. 
The formula is useful in discussing for example, the subbarrier 
enhancement of cross sections, 
providing reference cross sections in the absence of channel couplings. 
For relatively heavy systems, such as $^{16}$O+$^{144}$Sm, the formula 
indeed reproduces well the exact result except for the deep subbarrier region, 
where the parabolic approximation itself breaks down. On the other hand, 
for light systems, such as 
$^{12}$C+$^{12}$C, the Wong formula tends to overestimate the cross 
section. In this paper, we have extended the Wong formula by including the 
energy dependence of the parameters entering the formula, that is, the 
barrier height, barrier position, and the barrier curvature. Evaluating 
these parameters at the grazing angular momentum for each 
energy, rather than at $l=0$, we have shown that the energy-dependent 
version of Wong's formula reproduces the exact result well, even for light 
systems. 

The symmetrisation of the system leads to a Wong cross section possessing 
an oscillatory contribution, for which a compact formula can be derived 
based on the parabolic approximation. 
We have shown that the formula for the oscillatory contribution can also be 
extended to the energy-dependent version. 

Fusion oscillations are most significant in light symmetric systems with 
spin-zero nuclei. We have analyzed the experimental data for 
the $^{12}$C+$^{12}$C system and have argued that the fusion oscillations 
provide a strong constraint on the nuclear potential employed in a 
calculation. We have also analysed the 
$^{12}$C+$^{13}$C system, in which elastic neutron transfer again gives 
rise to oscillations. We have shown that these oscillations 
are useful in determining the sign of the effective parity-dependent potential 
arising from elastic transfer. 

We have argued that fusion oscillations provide an important tool 
for studying properties of the nuclear potential, strongly-coupled 
channels at high excitation energy, and fission. 
This is especially true for the $^{12}$C+$^{12}$C system, 
which plays an important 
role in several astrophysical phenomena and thus has been recognized as one of 
the key reactions~\cite{spillane,notani,jiang} in that domain.  
Understanding both the smooth and oscillatory parts of
the cross section above the barrier will almost certainly be necessary in 
understanding it in the important astrophysical region well below the barrier.

\begin{acknowledgments}
This work was supported by 
JSPS KAKENHI Grant Number 25105503. 
\end{acknowledgments}

\appendix

\section{Approximate formula for the energy-dependent barrier position}

For the exponential potential of Eq.~(\ref{eq:exp-pot}), 
the grazing angular momentum $l_g$ is given by 
\begin{eqnarray}
E&=&
-\frac{a\,Z_1Z_2e^2}{R_B^2}\,{\rm e}^{\delta R/a}+\frac{Z_1Z_2e^2}{R_E}+\frac{l_g(l_g+1)\hbar^2}{2mR_E^2}\nonumber \\
\end{eqnarray}
with $\delta R=R_B-R_E$, where $R_E$ is the position of the barrier for $l=l_g$ 
(see Eq.~(\ref{eq:graze})). Because the first derivative of the total potential 
is zero at the barrier position, we also have
\begin{eqnarray}
0&=&
\frac{Z_1Z_2e^2}{R_B^2}\,{\rm e}^{\delta R/a} -\frac{Z_1Z_2e^2}{R_E^2}-\frac{l_g(l_g+1)\hbar^2}{mR_E^3}.\nonumber \\
\end{eqnarray}
By combining these two equations, it can be shown that 
$\delta R$ satisfies 
\begin{equation}
{\rm e}^{\delta R/a}=1+\frac{E-B+\frac{1}{2}V_{CB}[1+\delta R/R_B-(1-\delta R/R_B)^{-1}]}{B-\frac{1}{2}V_{CB}(1+\delta R/R_B)},            
\label{eq:to-iterate}
\end{equation}
where $V_{CB}$ is again the Coulomb potential at $R_B$, the position of the $l=0$ barrier.

Inserting $\delta R=0$ on the right-hand side of this equation, one obtains 
\begin{eqnarray}
{\rm e}^{\delta R/a}=1+\frac{E-B}{B-\frac{1}{2}V_{CB}}
\end{eqnarray}
and from this first-order approximation to $\delta R$ we obtain (c.f. Ref.\cite{kabir-E-dep})  
\begin{eqnarray}
R_E=R_B-a\,{\rm ln}\left(1+\frac{E-B}{B-\frac{1}{2}V_{CB}}\right).
\label{eq:first}
\end{eqnarray}
Eq.~(\ref{eq:to-iterate}) can be easily iterated and rapidly converges to the 
exact result for $R_E$. 

\begin{figure}[tb]
\centering
\includegraphics[width=0.40\textwidth,angle=0,clip]{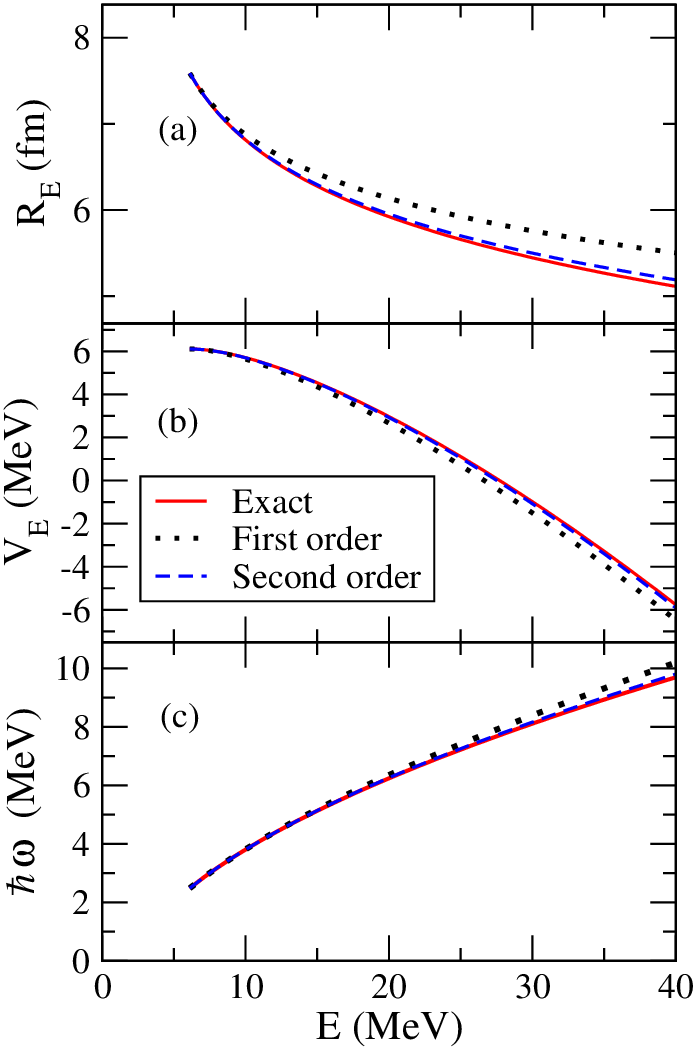}
\caption{(Color online) (a) exact and approximate 
barrier positions $R_E$ for the $^{12}$C+$^{12}$C system 
as a function of the incident energy $E$. 
An exponential nuclear potential with the diffuseness parameter $a$=0.8~fm 
has been used. The solid line 
shows the exact value obtained with Eq.~(\ref{eq:to-iterate}), while the  dotted
and the dashed lines are obtained from Eq.~(\ref{eq:first}) in first- and second-order
respectively. 
(b) as (a) but for the value of the $s$-wave potential at $R_E$. 
(c) same for the curvature of the barrier for the grazing angular 
momentum $l_g$.}  
\label{fig:rb2}
\end{figure}

The top panel of Fig.~\ref{fig:rb2} shows a comparison between 
the exact value of $R_E$ (solid line) 
with those obtained in first order (Eq.~(\ref{eq:first})) and second order
(dotted and dashed lines respectively)
for the $^{12}$C+$^{12}$C system. The middle and the bottom panels 
show $V_E$ and $\hbar\omega_E$ used in the energy-dependent  
Wong formula (\ref{eq:wong}), evaluated at the corresponding exact and approximate 
$R_E$ values. 
One can see that the first-order formula given by Eq.~(\ref{eq:first}) works well
only at energies in the vicinity of the $s$-wave Coulomb barrier, where $\delta R$ 
is small. The second-order result already leads to excellent results 
over the entire energy range shown, and higher orders coverge rapidly to the exact result.

\section{Derivation of oscillatory term}

In this appendix, we give a detailed derivation of Eqs.~(\ref{eq:poles}) 
and~(\ref{eq:osc}). 
First we note that the Hill-Wheeler formula for the transmission,
\begin{equation}
T(E,\lambda)=\frac{1}
{1+
\exp\left[\frac{2\pi}{\hbar\omega}
\left(B-E+\frac{\lambda^2\hbar^2}{2mR^2}\right)\right]},
\label{HW}
\end{equation}
has poles at the complex angular momenta $\lambda$ which satisfy 
\begin{equation}
\exp\left[\frac{2\pi}{\hbar\omega}
\left(B-E+\frac{\lambda^2\hbar^2}{2mR^2}\right)\right]=-1=e^{i(\pi+2n\pi)},
\label{pole}
\end{equation}
with $n=0,\pm 1,\pm 2,\cdots$. 
To find the poles nearest to the real axis with $n=0$ and $-1$, we put 
$\lambda_{\rm pole}=\lambda_g+i\lambda_I$, and neglecting the second-order term in 
$\lambda_I$, we find from Eq.~(\ref{pole}) that $\lambda_g$ and $\lambda_I$ 
satisfy
\begin{equation}
B-E+\frac{\lambda_g^2\hbar^2}{2mR^2}=0,
\end{equation}
and 
\begin{equation}
\lambda_I\sim \pm \frac{\hbar\omega}{2\lambda_g}\cdot\frac{mR^2}{\hbar^2}.
\label{impart}
\end{equation}
In the vicinity of the poles, the denominator of the right hand side of 
Eq.~(\ref{HW}) is given by 
\begin{eqnarray}
&&1+
\exp\left[\frac{2\pi}{\hbar\omega}
\left(B-E+\frac{\lambda^2\hbar^2}{2mR^2}\right)\right] \nonumber \\
&&\sim -\frac{2\pi}{\hbar\omega}\cdot\frac{\lambda_{\rm pole}\hbar^2}{mR^2}\,
(\lambda-\lambda_{\rm pole}),
\end{eqnarray}
where we have used Eq.~(\ref{pole}) to derive this equation. 

We are now ready to evaluate the $m=\pm 1$ terms in the Poisson summation 
formula (\ref{eq:poisson}) by contour integration. From the contour enclosing
the upper, right quadrant of the complex plane, we may write the $m=1$ term as
\begin{eqnarray}
\sigma_1&=&-\frac{2\pi}{k^2}\int^\infty_0\lambda T(E,\lambda)e^{2\pi i \lambda}d\lambda, \\
&=&-\frac{2\pi}{k^2}
\left[-2\pi i\lambda^{(+)}_{\rm pole}\frac{\hbar\omega}{2\pi}\frac{mR^2}
{\lambda_{\rm pole}^{(+)}\hbar^2}e^{2\pi i \lambda_{\rm pole}^{(+)}}\right. \nonumber \\
&&\left.-\int^0_\infty(i\tilde{\lambda})T(E,i\tilde{\lambda})e^{-2\pi\tilde{\lambda}}
(id\tilde{\lambda})\right],
\end{eqnarray}
where $\lambda^{(+)}_{\rm pole}$ takes the positive sign in Eq.~(\ref{impart}). 
Here, we neglect the second term of this equation; as $T$ is real and $<1$ on 
the imaginary axis, the integral is real and $<(2\pi)^{-2}$ and merely gives a very small 
correction to the smooth part of the cross section. 
So one finds 
\begin{equation}
\sigma_1
=i\frac{2\pi\hbar\omega}{k^2}\frac{mR^2}{\hbar^2}
\,e^{2\pi i \lambda_g}e^{-\pi\frac{\hbar\omega}{\lambda_g}\frac{mR^2}{\hbar^2}}.
\label{m1}
\end{equation}
Similarly for the $m=-1$ term we have
\begin{equation}
\sigma_{-1}
=i\frac{2\pi\hbar\omega}{k^2}\frac{mR^2}{\hbar^2}
\,e^{-2\pi i \lambda_g}e^{-\pi\frac{\hbar\omega}{\lambda_g}\frac{mR^2}{\hbar^2}}.
\label{mm1}
\end{equation}
Combining Eqs. (\ref{m1}) and (\ref{mm1}) and using 
$\sin(2\pi\lambda_g)=\sin(2\pi l_g+\pi)=-\sin(2\pi l_g)$, 
one finally obtains Eq.~(\ref{eq:poles}). 

With identical particles the fusion cross sections must be symmetrized
and are given by
\begin{eqnarray}
\sigma&=&\frac{\pi}{k^2}\sum_l(2l+1)T_l(E)(1\pm(-1)^l), 
\label{fussym} \\
&=&
\frac{2\pi}{k^2}\sum_{m=-\infty}^{m=\infty}
(-1)^m\int^\infty_0\lambda d\lambda T(E,\lambda)e^{2\pi m i\lambda}
(1\mp ie^{i\pi \lambda}), \nonumber \\
\end{eqnarray}
where the positive sign in Eq.~(\ref{fussym}) relates to the spatially 
symmetric case and the negative sign to the spatially 
anti-symmetric case. The symmetrization is seen to lead to two terms, $m=0$ and $m=-1$,
where the exponent is reduced by a factor 2. These terms now clearly dominate,
and the fusion cross sections are approximately given by 
\begin{eqnarray}
\sigma
&=&
\frac{2\pi}{k^2}
\int^\infty_0\lambda d\lambda T(E,\lambda)
(1\mp ie^{i\pi \lambda}) \nonumber \\
&&\pm 
\frac{2\pi}{k^2}
\int^\infty_0\lambda d\lambda T(E,\lambda)e^{-i\pi \lambda}. 
\end{eqnarray}
One can evaluate these integrals as above, 
and using $\cos(\pi\lambda_g)=-\sin(\pi l_g)$, we obtain 
\begin{equation}
\sigma=\sigma_{\rm Wong}\pm\sigma_{\rm osc},
\end{equation}
where the considerably enhanced oscillatory term is now given by Eq.~(\ref{eq:osc}).

In the above proof, we have used energy-independent values for $B$,
$R$ and $\hbar \omega$ for simplicity of notation. However, the results are
easily generalised to the energy-dependent case, when the above total cross section is 
given by Eqs.~(\ref{eq:wong}) and (\ref{eq:osc}) using $V_E$, $R_E$ and $\hbar \omega_E$. 
(More generally these quantities depend on the angular momentum rather 
than the energy, see discussion in Sec.~III).

\end{document}